\documentclass[12pt]{iopart}
\usepackage{iopams}
\usepackage{graphicx,wrapfig,epsfig}
\begin{document}

\title{Construction and Expected Performance of the Hadron Blind Detector for the PHENIX Experiment at RHIC}

\author{A Milov$^1$,
W~Anderson$^4$,
B~Azmoun$^1$,
C-Y~Chi$^2$,
A~Drees$^3$,
A~Dubey$^4$,
M~Durham$^3$,
Z~Fraenkel$^4$,
J~Harder$^1$,
T~Hemmick$^3$,
R~Hutter$^3$,
B~Jacak$^3$,
J~Kamin$^3$,
A~Kozlov$^4$, 
M~Naglis$^4$,
P~O'Connor$^1$,
R~Pisani$^1$,
V~Radeka$^1$,
I~Ravinovich$^4$,
T~Sakaguchi$^1$,
D~Sharma$^4$,
A~Sickles$^1$,
S~Stoll$^1$,
I~Tserruya$^4$,
B~Yu$^1$,
C~Woody$^1$
}

\address{$^1$ Brookhaven National Laboratory, Upton, NY 11973-5000, USA}
\address{$^2$ Nevis Columbia Laboratories, Irvington, NY 10533, USA}
\address{$^3$ Stony Brook University, SUNY, Stony Brook, NY 11794, USA}
\address{$^4$ Weizmann Institute, Rehovot 76100, Israel}
\ead{amilov@bnl.gov}

\begin{abstract}
A new Hadron Blind Detector (HBD) for electron identification in high density hadron
environment has been installed in the PHENIX detector at RHIC in the
fall of 2006. The HBD will 
identify low momentum electron-positron pairs to
reduce the combinatorial background  in the $e^{+}e^{-}$ mass
spectrum, mainly in the low-mass region below 1
GeV/c$^{2}$. The HBD is a windowless proximity-focusing Cherenkov detector with a 
radiator length of 50 cm, a CsI photocathode and three layers of Gas
Electron Multipliers (GEM). The HBD uses pure CF$_{4}$ as a radiator
and a detector gas. Construction details and the expected performance
of the detector are described.
\end{abstract}



\section{Physics goal and detector concept}
One of the primary goals of the PHENIX~\cite{phenix} experiment at RHIC is to study
the production of low-mass electron-positron pairs~\cite{phenix_dilept} as
a tool to investigate the properties of the new state of matter
discovered at RHIC. Results of similar studies published by CERN
experiments~\cite{na45,na60} at lower energy show an excess of $e^{+}e^{-}$ pairs
produced in the mass region below 1~GeV/c$^2$. The main difficulty to a
measurement in this mass region is the combinatorial background from
the $e^+e^-$ pairs with a small opening angle coming from
the $\pi^{0}$-meson decays and $\gamma$-conversions. To overcome this problem the PHENIX
experiment designed and built a new detector to reject such $e^+e^-$ pairs.

The detector concept is described in~\cite{loi}. The HBD is a proximity-focusing windowless Cherenkov counter. Pure CF$_{4}$ is used as the
radiator gas and the amplification media. This choice
has several advantages crucial for the detector operation. A high
refraction index (n=1.00062 in the visible region and higher in the UV)
allows production of more Cherenkov photons than most other gases. CF$_{4}$ is fully
transparent up to 11.5 eV which makes it a very good match to the CsI
photocathode with $>$70\% quantum efficiency in the deep UV. The
electron extraction efficiency into CF$_{4}$ is among the highest of
any gas; e.g. more than a factor of 2 higher than that into $Ar$. 
CF$_{4}$ based mixtures are used as working gas in many detectors, including
PHENIX. We have shown~\cite{sasha,RandD} that pure
CF$_{4}$ can be used as a detector gas. 

\begin{figure}[h]
  \includegraphics[height=.23\textheight]{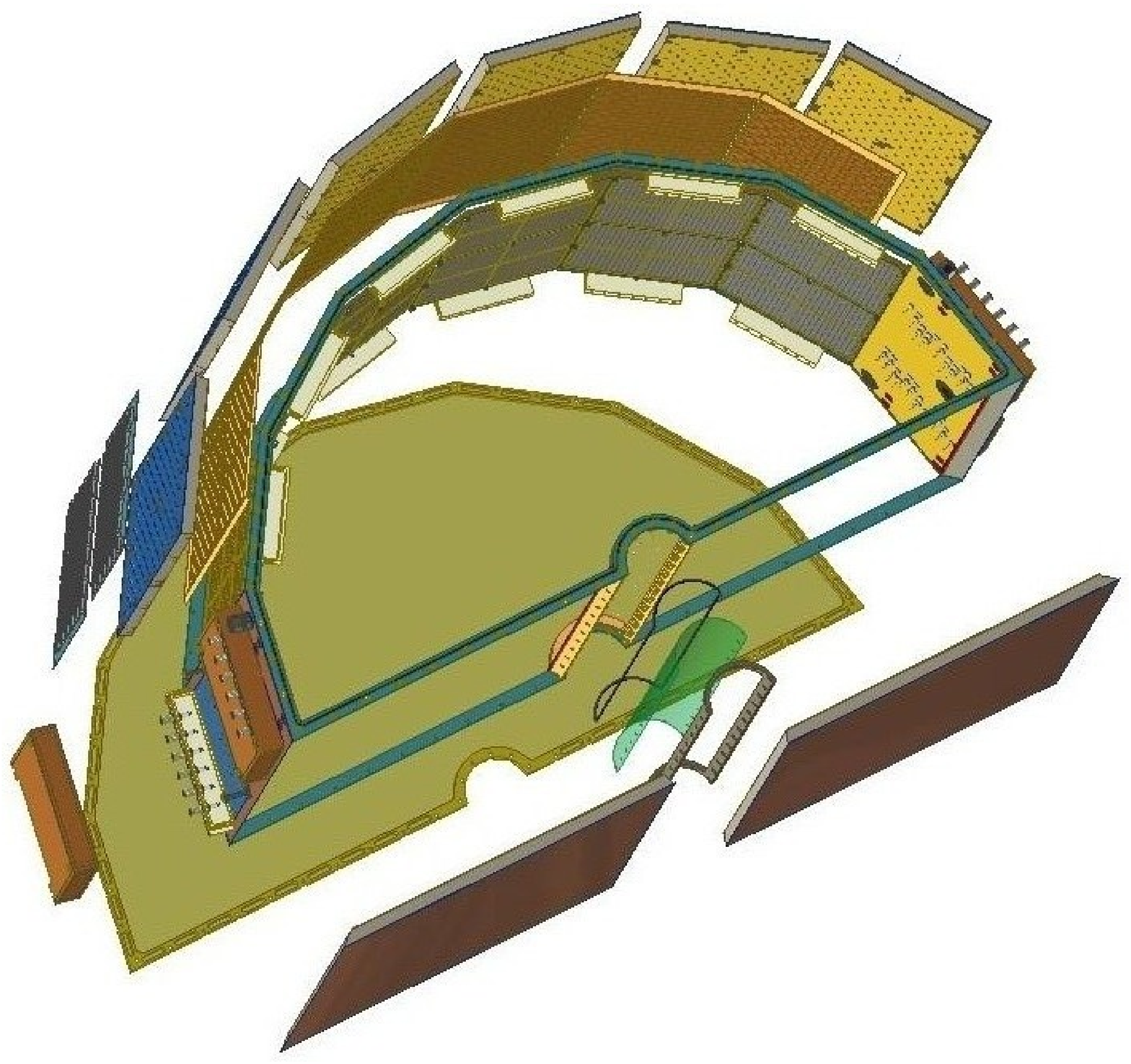} \hspace{1cm}
  \includegraphics[height=.20\textheight]{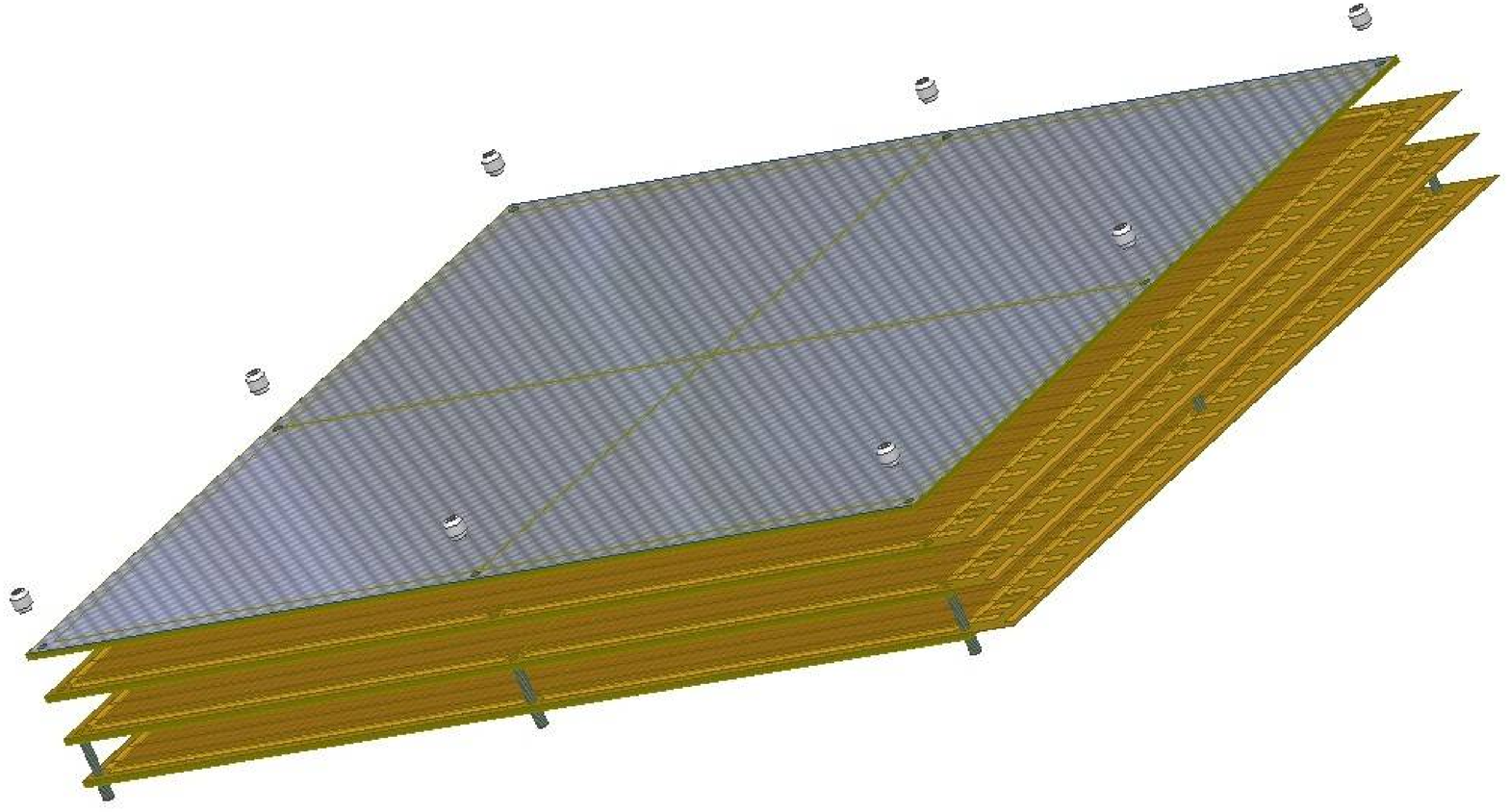}
  \caption{Left: Exploded view of the HBD detector. One side panel is
removed for clarity. Right: A photodetector module
consisting of 3 GEMs and a mesh.\label{fig:gems}}
\end{figure}

The HBD is located close to the interaction vertex. It starts after
the beam pipe at a radius of 5~cm and extends up to an outer radius of
65~cm. The HBD is made of two identical arms. An exploded view of one
detector arm is shown in Fig.~\ref{fig:gems}.
A thin FR4 frame holds panels made of FR4/Honeycomb/FR4 (250$~\mu$m/19~mm/250~$\mu$m) sandwiches. The
2 larger front panels and the 8 smaller back panels are glued to a
frame that provides rigidity to the box. The two side panels (the closer one is removed for clarity) are mounted with plastic screws. Among the 8 back panels the
outer 2 are reserved for the detector services such as gas in/out,
high voltage, UV
windows and the 6 inner panels are equipped each with 2 photodetector modules on
the inside and connected to the Front End Electronics (FEE) board attached to the
outer surface of the detector. Electrons produced in the collision
enter the detector through a 127~$\mu$m mylar window coated with 100~nm
of Al and radiate
Cherenkov photons along their path in 50~cm of CF$_{4}$ radiator.

The photodetector shown in the right panel of Fig.~\ref{fig:gems} is a
stack of 3 GEMs 27$\times$22~cm$^2$ stretched on a 1.5~mm thick and
5~mm wide FR4 frame. A cross-like 0.3~mm wide support runs in the middle of the
frame. The top GEM facing the detector volume has a 0.2-0.4~$\mu$m layer of CsI
evaporated on its surface also plated with gold to prevent CsI
poisoning by the Cu of the GEM. A 90\% transparent stainless steel mesh
1.5~mm above the stack is biased by a positive voltage such that
the ionization from charged particles drifts away from the GEM holes
to keep the detector insensitive (i.e. ``blind'') to signals from charged
hadrons. Details about HBD photodetectors can be found in~\cite{loi,sasha,it,sam}.

The inner surface of the panels is covered by a single Kapton film
with 1152 hexagonal pads printed on it. This film also serves as an additional gas
seal. Each pad has an area of 6.3 cm$^{2}$ and is connected to a charge
sensitive amplifier on the FEE board. The size of the pad is slightly smaller than the
spread of the Cherenkov photons emitted by a single particle and
larger than the size of a single avalanche in the GEM. It makes
the Cherenkov signal appear in several adjacent pads,
while an ionizing particle produces signal primarily in a single pad.

\section{Full scale prototype}
A full scale prototype of the HBD with one instrumented sector was
installed in PHENIX during the $p+p$ physics run in
2006~\cite{craig}. The magnetic field in the PHENIX was switched off and a
special trigger was set up to increase the electron sample in the
active area.
\begin{figure}[h]
  \includegraphics[height=.19\textheight]{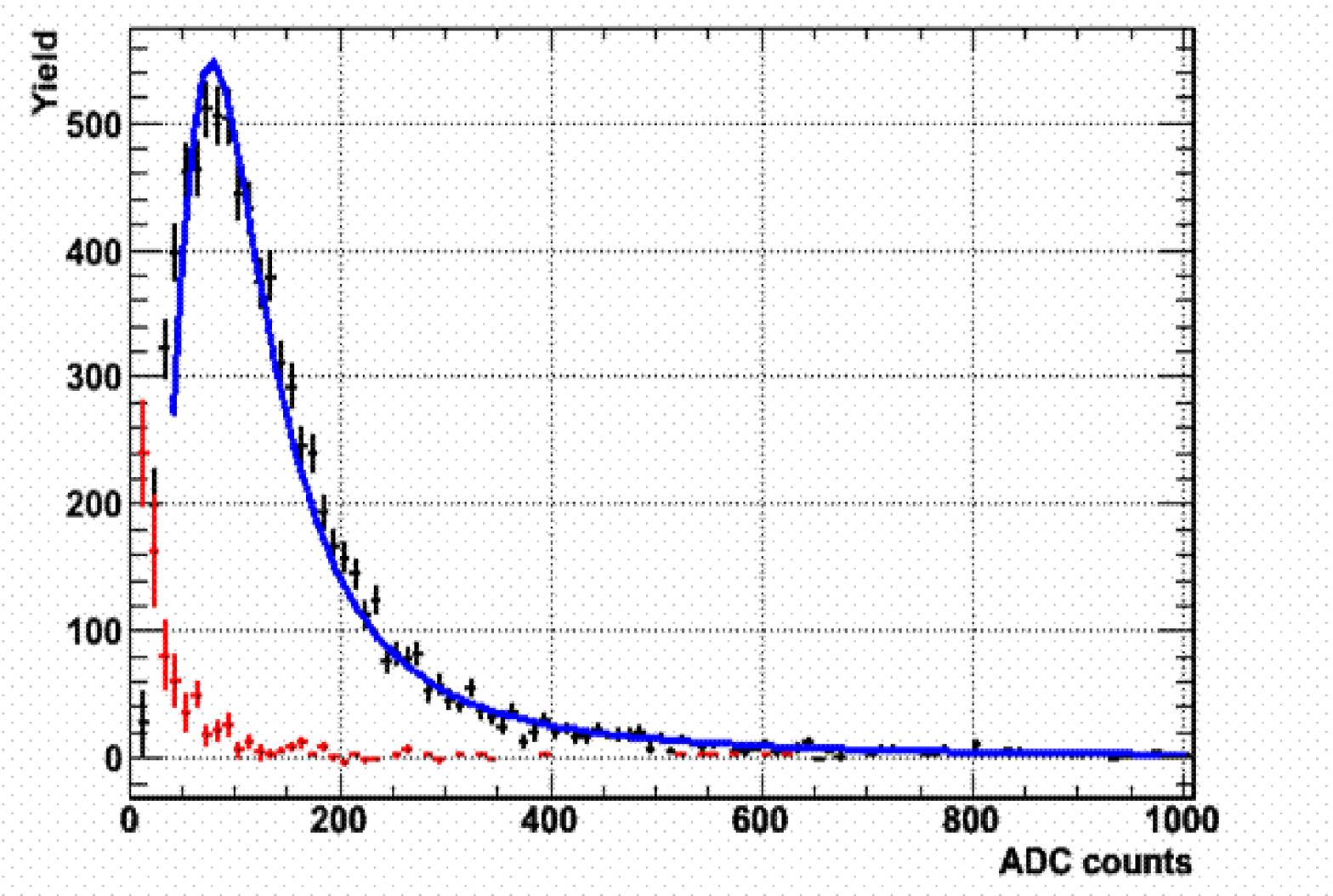} \hspace{1cm}
  \includegraphics[height=.19\textheight]{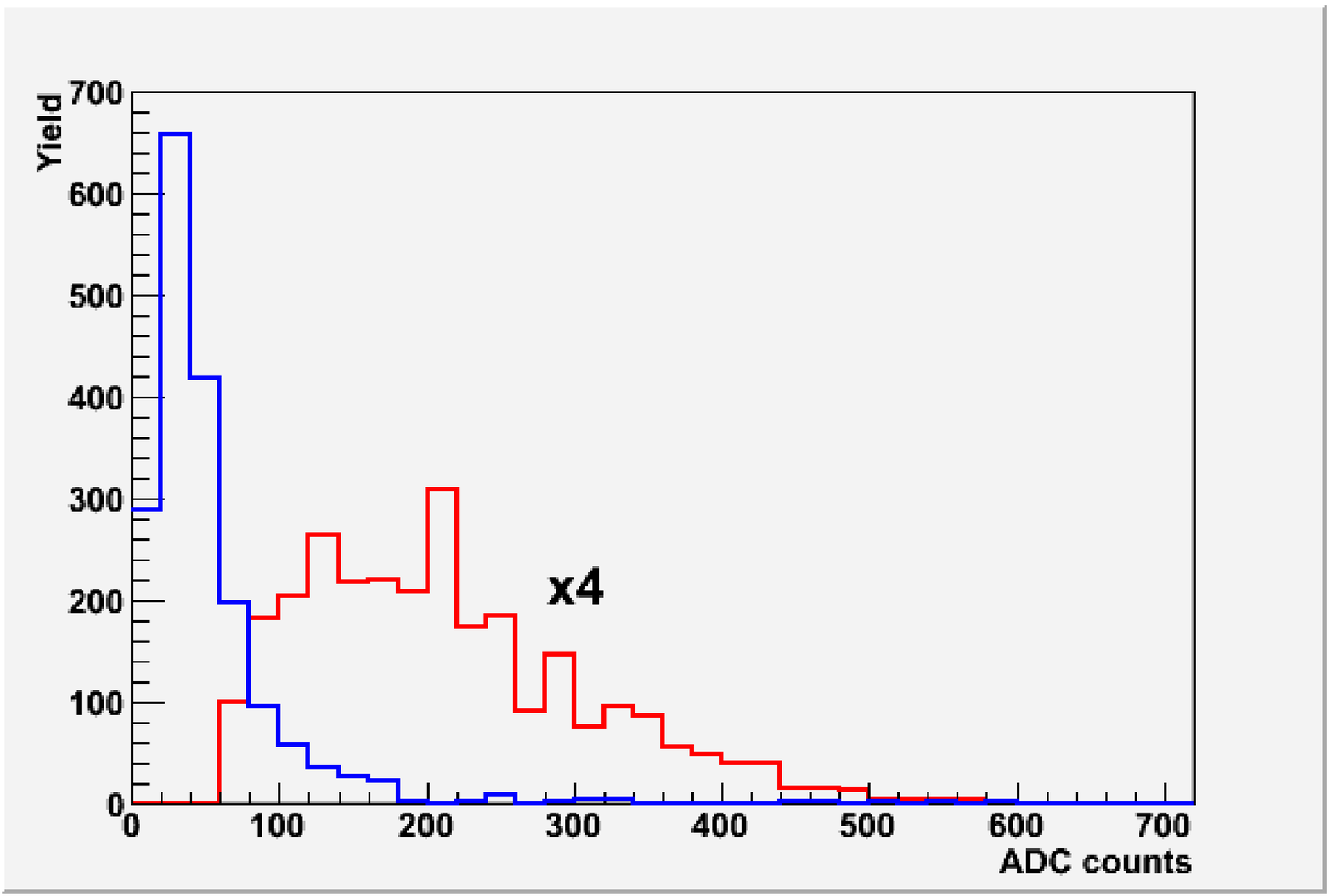}
  \caption{Left: The cluster amplitude measured with the forward (collecting)
voltage bias are shown in blue (color
online) and reversed bias in red. Right: all tracks (blue) and
electrons (red) measured in the reversed bias configuration.\label{fig:bias}}
\end{figure}

The cluster amplitude distribution produced by all particles measured with
a forward (collecting) voltage bias between the mesh and the GEM is
shown in the left panel of the Fig.~\ref{fig:bias}. The distribution
follows the Landau shape characteristic of minimum ionizing particle
signals. The same distribution measured in the reversed bias configuration
has much lower amplitudes and is shown with red points. In the right panel the same distribution measured with the reversed
bias configuration is compared to a distribution produced by particles
which are identified as electrons by the PHENIX electron
identification. Signals from ionizing particles
have significantly lower amplitudes than signals from
electrons shown in red. Electrons can be separated from hadrons by a cut at around
100 ADC channels. The average cluster size for hadrons and electrons is also different. It was
found to be 1.2 pads and 3 pads respectively.

\section{The final detector}
Both arms of the HBD were installed in PHENIX by the end of
2006. For the construction of the photodetectors 133 GEMs (85 standard
plus 48 gold plated) were used. 65+37 GEMs passed the initial tests. These GEMs were measured for gas gain
uniformity and 48+24 were selected and combined to
minimize the gas gain variation across the surface of the triple GEM
stacks. The gas gain uniformity was measured to be between 5\% and 20\% in all 24
modules. The CsI photocathodes were simultaneously evaporated on 4 gold-plated GEMs
using the evaporator built by the INFN group and the group of Istituto
Superiori di Sanita, Rome~\cite{infn}. The quantum efficiency was
measured and found to be the same as in the R\&D
studies~\cite{sasha,it,ir} and uniform
within better than 5\% across the area of the GEM.

\begin{wrapfigure}{l}{10cm}
  \includegraphics[height=.19\textheight]{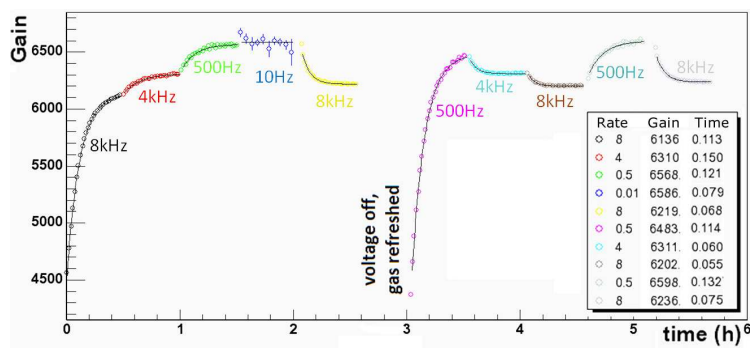}
  \caption{Gain of the GEM stack exposed to
different rates of the $^{55}$Fe source vs. time.\label{fig:gain}}
\end{wrapfigure}

During tests of the GEM stack with $^{55}$Fe source we measured the
dependence of the gas gain on time and rate as shown in Fig.~\ref{fig:gain} for different exposure rates.
The HBD operation conditions in Au-Au run of RHIC are equivalent to
a source rate below 100~s$^{-1}$ which impose no problem to
the detector operation. However, the initial increase of gain with time may
require more than one hour before stable gain conditions are
established. The GEM stacks have to be under 
the operation voltage
well before the data taking begins.

The authors acknowledge support from the Department of Energy, NSF
(U.S.A.) Israel Science Foundation, US-Israel BSF and the Nella and
Leon Benoziyo Center of High Energy Physics Research. The work of one of
us (A.M.) is supported by the
Goldhaber Fellowship at BNL with funds provided by Brookhaven
Science Associates. 

\end{document}